\documentclass[traditabstract]{aa}
\usepackage{aas_macros}

\usepackage{amsmath}
\usepackage{rotating}
\usepackage{float}
\usepackage{graphicx}	
\usepackage{txfonts}
\usepackage{natbib} 
\usepackage{xspace}
\usepackage{url}
\bibpunct{(}{)}{;}{a}{}{,} 
\usepackage{subfigure}

\begin{document}
\title{A PCA-based automated finder for galaxy-scale strong lenses}
\titlerunning{PCA-based lens finder}

\author{R. Joseph\inst{1} \and F. Courbin\inst{1}  \and
R.B. Metcalf \inst{2} \and 
C. Giocoli \inst{2,3,4} \and
P. Hartley \inst{5} \and 
N. Jackson \inst{5}\and
F. Bellagamba \inst{2} \and
J.-P. Kneib \inst{1} \and
L. Koopmans \inst{6} \and
G. Lemson \inst{7}
M. Meneghetti \inst{3,4,8} \and
G. Meylan \inst{1} \and
M. Petkova \inst{2,9,10} \and
S. Pires \inst{11}
}
\authorrunning{R. Joseph et al.}

\institute{
Laboratoire d'astrophysique, Ecole Polytechnique F\'ed\'erale de Lausanne (EPFL), Observatoire de Sauverny, CH-1290 Versoix, Switzerland \and  
Dipartimento di Fisica e Astronomia - Universita di Bologna, via Berti Pichat 6/2, I-40127 Bologna, Italy \and 
INAF - Osservatorio Astronomico di Bologna, via Ranzani 1, 40127, Bologna, Italy \and  
INFN - Sezione di Bologna, viale Berti Pichat 6/2, 40127, Bologna, Italy \and 
Jodrell Bank Centre for Astrophysics, School of Physics $\&$ Astronomy, University of Manchester, Oxford Road, Manchester M13 9PL, UK \and 
Kapteyn Astronomical Institute, University of Groningen, PO Box 800, NL-9700 AV Groningen, the Netherlands \and
Department of Physics, Ludwig-Maximilians-Universit\"at, Scheinerstr. 1, D-81679 M\"unchen, Germany \and 
Jet Propulsion Laboratory, 4800 Oak Grove Dr., La Canada-Flintridge, CA 91011, USA \and
Max-Planck-Institut f\"ur Astrophysik, D-85748 Garching, Germany \and 
Excellence Cluster Universe, Boltzmannstr. 2, D-85748 Garching, Germany \and
Laboratoire AIM, CEA/DSM-CNRS-Universite Paris Diderot, IRFU/SEDI-SAP, Service d'Astrophysique, CEA Saclay, Orme des Merisiers, 91191 Gif-sur-Yvette, France
}

\date{Received ; accepted }


\abstract{We present an algorithm using Principal Component Analysis (PCA) to subtract galaxies from imaging data, and also two algorithms to find strong, galaxy-scale gravitational lenses in the resulting residual image. The combined method is optimized to find full or partial Einstein rings. Starting from a pre-selection of potential massive galaxies, we first perform a PCA to build a set of basis vectors. The galaxy images are reconstructed using the PCA basis and subtracted from the data. We then filter the residual image with two different methods. The first uses a curvelet (curved wavelets) filter of the residual images to enhance any curved/ring feature. The resulting image is transformed in polar coordinates, centered on the lens galaxy center. In these coordinates, a ring is turned into a line, allowing us to detect very faint rings by taking advantage of the integrated signal-to-noise in the ring (a line in polar coordinates). The second way of analysing the PCA-subtracted images identifies structures in the residual images and assesses whether they are lensed images according to their orientation, multiplicity and elongation.  We apply the two methods to a sample of simulated Einstein rings, as they would be observed with the ESA Euclid satellite in the VIS band. The polar coordinates transform allows us to reach a completeness of 90\% and a purity of 86\%, as soon as the signal-to-noise integrated in the ring is higher than 30, and almost independent of the size of the Einstein ring. Finally, we show with real data that our PCA-based galaxy subtraction scheme performs better than traditional subtraction based on model fitting to the data. Our algorithm can be developed and improved further using machine learning and dictionary learning methods, which would extend the capabilities of the method to more complex and diverse galaxy shapes.}


\maketitle

\section{Introduction}

With the many ongoing or planned sky surveys taking place in the optical and near-IR, gravitational lensing has become a major scientific tool to study the properties of massive structures at all spatial scales. On the largest scales, in the weak regime, gravitational lensing constitutes a crucial cosmological probe \citep[e.g.][]{Heymans2013, Frieman2008}. On smaller scales, weak galaxy-galaxy lensing allows us to study the extended halo of individual galaxies or of groups of galaxies \citep[e.g.][]{Simon2012} and to constrain cosmology \citep[e.g.][]{Mandelbaum2013, Parker2007}. 

In the strong regime, when multiple images of a lensed source are seen, gravitational lensing offers an accurate way to weigh galaxy clusters  \citep[][for reviews]{Bartelmann2013, Hoekstra2013, Meneghetti2013, Kneib2011}, galaxy groups \citep[e.g.][]{Foex2013, Limousin2009} and individual galaxies \citep[e.g.][]{Brownstein2012, Treu2011, Bolton2006}. However, all strongly lensed systems known today, combined together, represent only hundreds of objects. Wide field surveys have the potential to produce samples three orders of magnitude larger, allowing us to study statistically dark matter and its evolution in galaxies as a function, e.g. of morphological type, mass, stellar and gas contents \citep[see][]{Gavazzi2012, Ruff2011, Sonnenfeld2013, SonnenfeldGavazzi2013}. For example, \citet{Pawase2012} predicts that a survey like Euclid will find at least 60000 galaxy-scale strong lenses. To find and to use them efficiently, it is vital to devise automated finders that can produce samples of lenses with high completeness and purity and with a well defined selection function.  Note that the lenses of \citet{Pawase2012} are source selected. There is no volume-limited sample of lens-selected systems, so the number 60000 systems is given here only to give an order of magnitude of the number of objects that future wide-field surveys will have to deal with.

\begin{figure*}[t!]
\begin{center}
{\includegraphics[width=4.0cm, height=4.0cm]{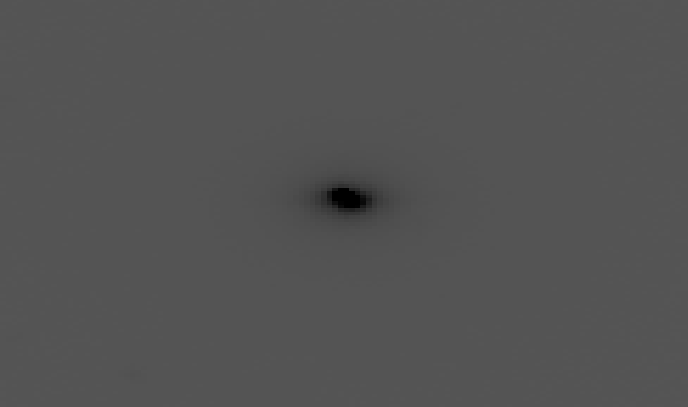}}
{\includegraphics[width=4.0cm, height=4.0cm]{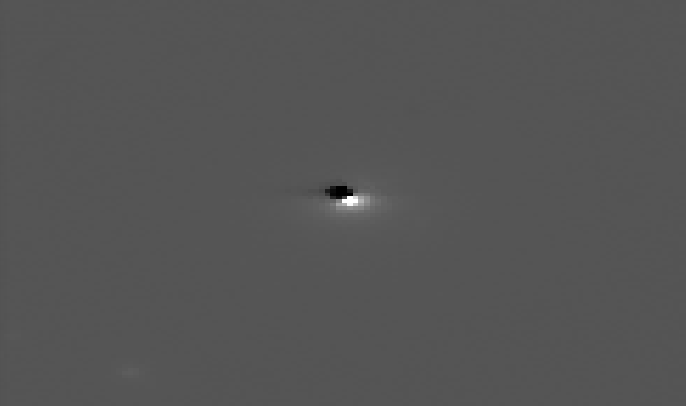}}
{\includegraphics[width=4.0cm, height=4.0cm]{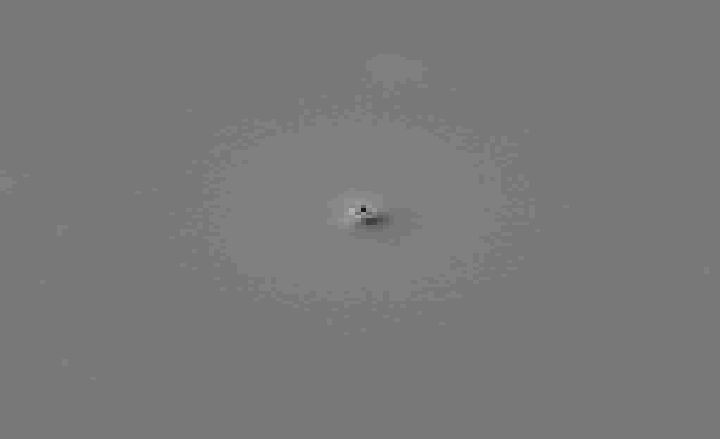}}
{\includegraphics[width=4.0cm, height=4.0cm]{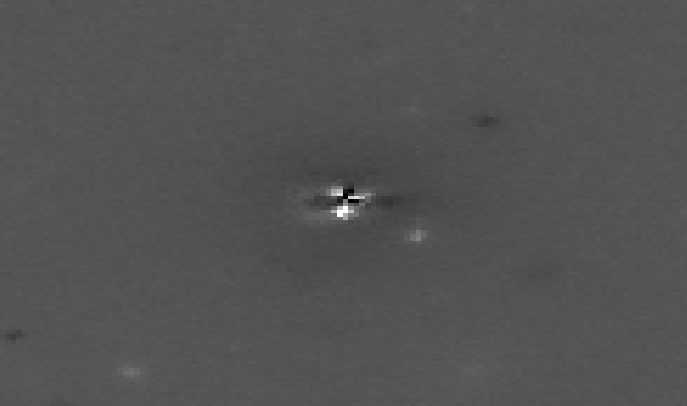}}
\caption{Examples of PCA components obtained using 1000 simulated galaxies from the Bologna Lens Factory (see Sect.~\ref{Euclid}).\label{fig:PCA_coefs}}
\end{center}
\end{figure*}

Several automated robots exist to find strong lenses. Among the best ones are {\tt Arcfinder} \citep{Seidel2007}, which was primarily developed to find large arcs behind clusters and groups, and the algorithm by \citet{Alard2006} used by \citet{Cabanac2007} and \citet{More2012}, to look for arcs produced by individual galaxies and groups in the CFHT Strong Lensing Legacy Survey. Other automated robots consider any galaxy as a potential lens and predict where lensed images of a background source should be before trying to identify them on the real data \citep{Marshall2009}. In order to detect lenses with small Einstein radii or with faint rings, most of these algorithms rely on foreground lens subtraction \citep[e.g.][]{Gavazzi2012}. So far, this subtraction has been performed through model fitting. 
 An example of a ring detector is given in \citet{Sygnet2010} which selects objects with possible lensing configuration according to their lensing convergence, estimated from the Tully-Fisher relation. This algorithm relies on photometric information but requires a visual check of a large number of candidates.

In the present paper, we propose a "lens finder" which uses single-band images to find full and partial Einstein rings based on purely morphological criteria. The algorithm uses as input a pre-selection of potential lens galaxies, hence producing so-called "lens-selected" samples. The present work sets the basis of an algorithm using machine learning techniques. Although focused on finding Einstein rings, it can be adapted to other types of lenses, such as those consisting of multiple, relatively pointlike, components.

This paper is organised as follows. In Sections \ref{PCA} and \ref{Finder} we outline our algorithm and introduce the principles behind each step of the process. In section \ref{Euclid} we show the performance of our algorithm using a set of simulations designed to reproduce Euclid images in the optical. We discuss the completeness and purity of our algorithm as a function of signal-to-noise (SNR) and caustic radius of the lensing systems. Section \ref{Stripe82} shows results of our galaxy subtraction algorithm compared to those of {\tt galfit} software \citep[][]{Peng2011} on images from the CFHT optical imaging of SDSS stripe 82 and Section \ref{conclusion} summarizes our main results.

\section{A new automated lens finder}
\label{PCA}

\subsection{Principle of the algorithm}

By construction, lens-selected samples display bright foreground lenses and faint background sources, otherwise the pre-selection of the lenses based on morphological type, luminosity and color would not be possible. As a consequence, faint Einstein rings are hidden in the glare of the foreground lenses, which must be properly removed before any search for lensed rings can be undertaken. An efficient "lens finder" therefore involves two main steps: 1- removal of the lens galaxy, 2- identification of rings in the lens-subtracted image.

A traditional way of subtracting galaxies is to fit a two dimensional elliptical profile to the data, e.g. as done with the {\tt galfit} software \citep[][]{Peng2011}. While this is sufficient to characterize the main morphological properties of galaxies, it turns out to be insufficient to detect faint arcs seen superposed on bright galaxies with a significant level of resolved structures.

One way to circumvent the problem is to build an empirical light model from the sample of galaxies itself, i.e. to use machine learning techniques such as Principal Component Analysis \citep[PCA;][]{Jolliffe1986}.  The sparsity and the diversity in terms of shape of the {\it  lensed objects} (rings, arcs, multiple images) prevents them from being well enough represented in the basis, hence allowing for an accurate separation of lenses and sources. This has already been used to find lensed sources from PCA decomposition of quasar spectra \citep[e.g.][]{Courbin2012, Boroson2010}. We adopt now a similar strategy to analyse images. 

Once the foreground lenses have been properly removed, we analyse the residual rings using methods described in Section ~\ref{Finder}. The main steps of the algorithm can be summarized as follows:

\begin{enumerate}
\item Pre-selection of the galaxies with a predefined range of shape parameters (size, ellipticities, magnitudes, colors, etc.)
\item Building the PCA basis either from the selected sample of galaxies or from an adapted training set.
\item Reconstruction of the central galaxies and subtraction from the original images. 
\item Detection of lensed objects, either using island finding (groups of adjacent pixels) or a polar transform or the resi\-dual image.
\end{enumerate}


\begin{figure*}[t!]
\begin{center}
\includegraphics[width=3.7cm, height=3.7cm]{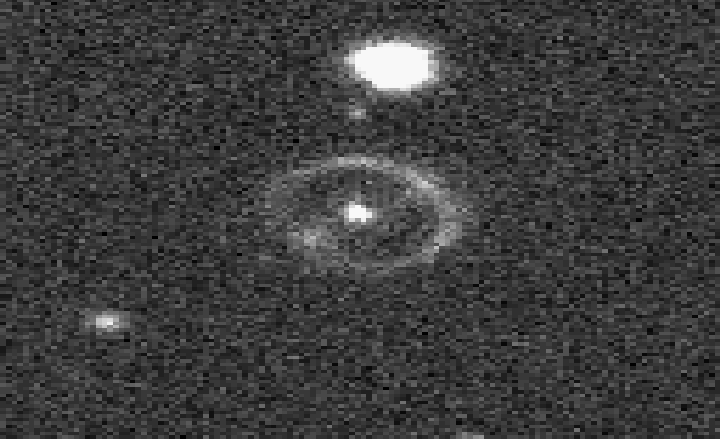}
\includegraphics[width=3.7cm, height=3.7cm]{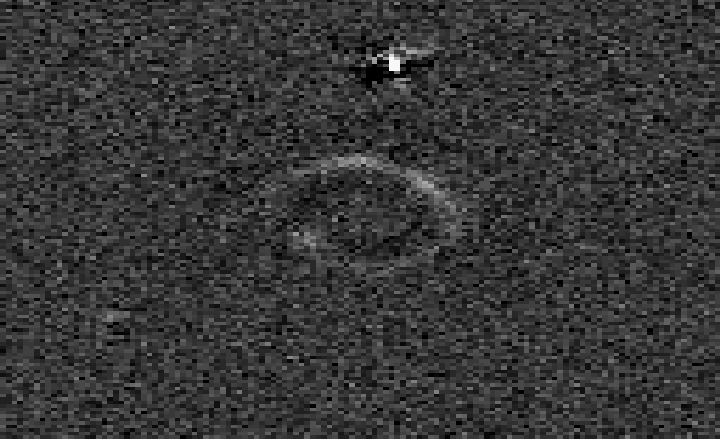}
\includegraphics[width=3.7cm, height=3.7cm]{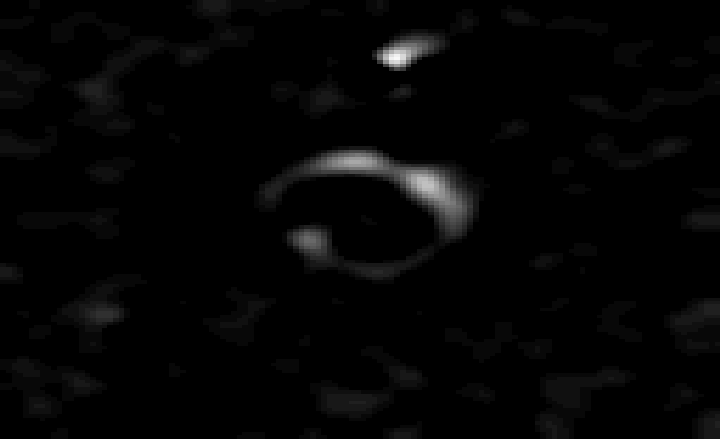}
\includegraphics[width=3.7cm, height=3.7cm, angle=90]{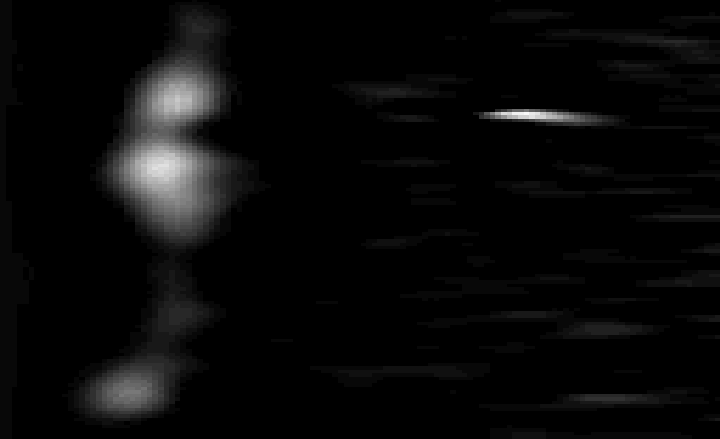}
\includegraphics[width=3.7cm, height=3.7cm]{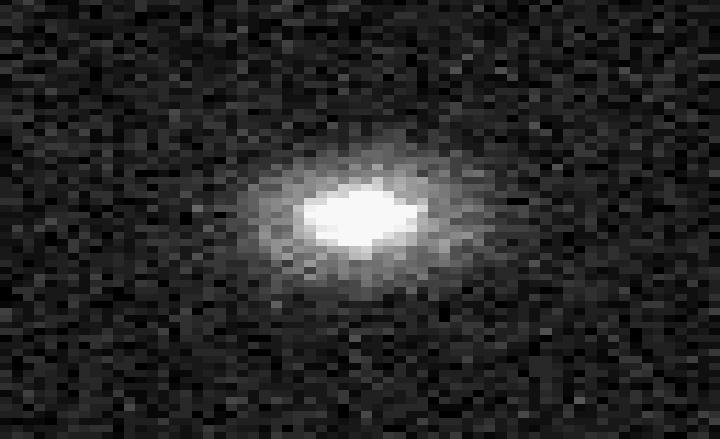}
\includegraphics[width=3.7cm, height=3.7cm]{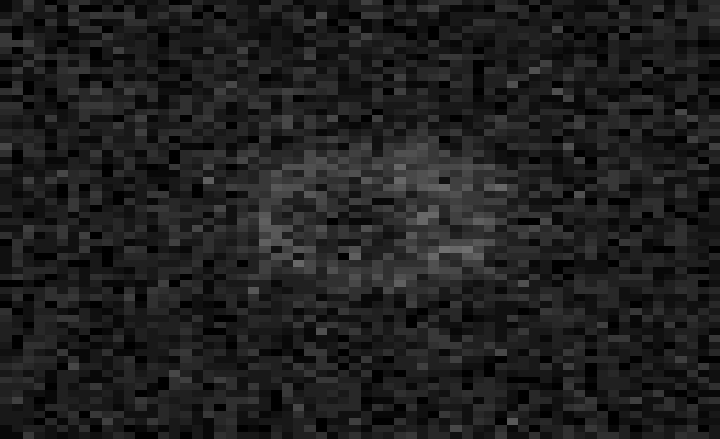}
\includegraphics[width=3.7cm, height=3.7cm]{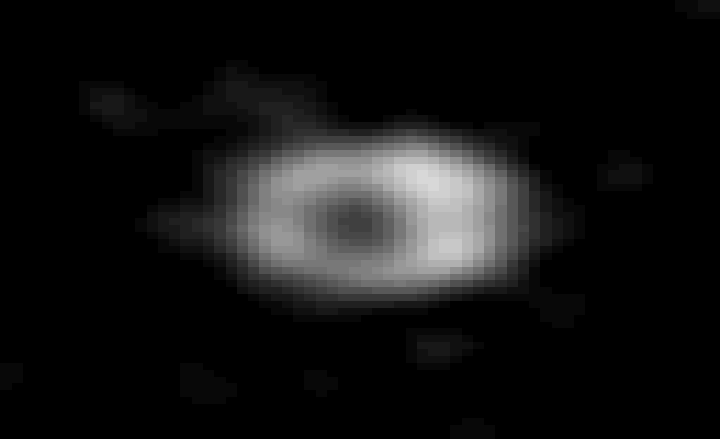}
\includegraphics[width=3.7cm, height=3.7cm, angle=90]{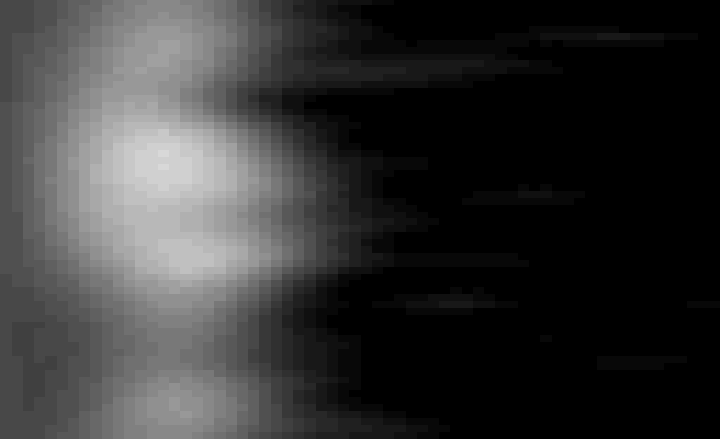}
\caption{Illustration of the ring finding process for two simulated Einstein rings from the Bologna Lens Factory (Sect.~\ref{Euclid}). For each row, from left to right are shown 1- an example of simulated Einstein ring (64$\times$64 pixels), along with its lens galaxy, 2- the lensed ring after PCA subtraction of the foreground galaxy, 3- the result of curvelet denoising, 4- the polar transform of the ring revealing a well visible horizontal line which position along the y-axis gives a measurement of the radius of the Einstein ring. \label{fig:lens_detector}}
\end{center}
\end{figure*}

\begin{figure}[t!]
\begin{center}
\begin{tabular}{c}
\includegraphics[trim=0cm 0.9cm 0cm 0cm, clip=true, width=8.5cm]{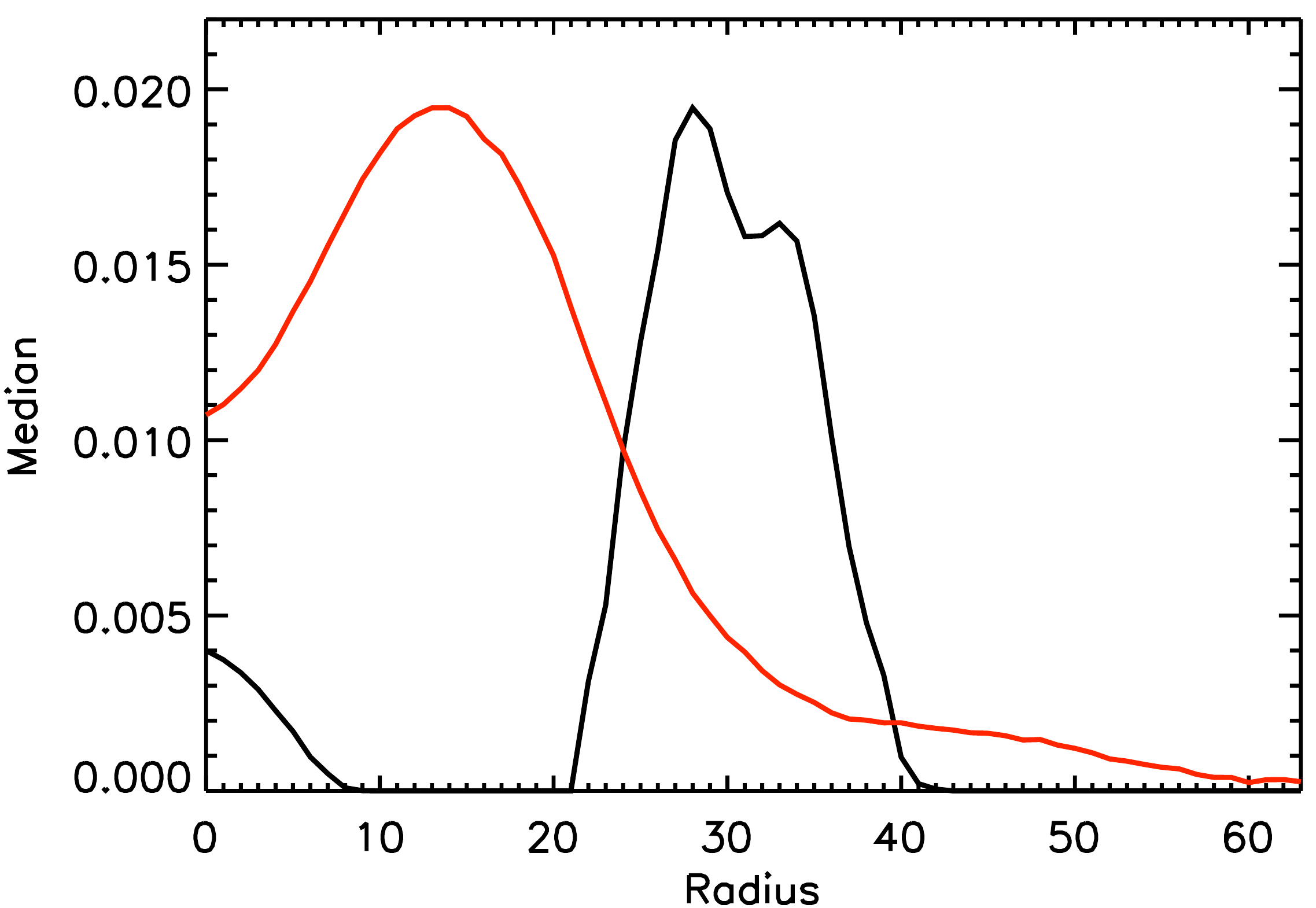}\\
{\tt Radius (in pixels)}
\end{tabular}
\caption{Median pixel values along the pixel rows of the curvelet-filtered images shown in the third column of Fig.~\ref{fig:lens_detector}. The black line corresponds to the top row of Fig.~\ref{fig:lens_detector} and the red line corresponds to the bottom row. A simple thresholding scheme allows us to detect the spike and to measure directly the size of the Einstein ring (see text).\label{fig:spike} }
\end{center}
\end{figure}

\subsection{Selection of galaxies}

The first step of this method is to build stamp images of galaxies in which to look for lensed objects. This step strongly depends on the specific sample considered and may take advantage of algorithms such as {\tt SExtractor} \citep{Bertin1996}.

For the PCA decomposition to work well, a compromise has to be found between the number of objects used to build the PCA basis, the size of the objects in pixels, and the range in shape parameters. The more complex the galaxies are, the more galaxies should be included in the training set, i.e. the sparsity of the problem has to be evaluated carefully. 

For relatively simple galaxy shapes, like elliptical galaxies, the pre-selection may focus on galaxies with similar sizes and ellipticities, which ensures better morphological similarities. This usually results in a satisfactory subtraction of the lens galaxy with only few PCA components. However, the window in which the sizes and ellipticities are chosen has to be wide enough to allow a full representation of any shapes of galaxies in this range. The choice of this selection window is discussed later when applying the method to specific data.

Computational time is an important parameter to consider as well. Building the PCA basis involves finding the eigenvectors and the eigenvalues of a $n^{2}\times N_{\rm gal}$ matrix, where $n$ is the number of  pixels per stamp and where $N_{\rm gal}$ is the number of stamps in the training set.

\subsection{Building the PCA basis}

Before computing the PCA basis, we rotate all the galaxies in the training set so that their major axes are all aligned and we center the galaxies in each stamp image. The rotation is performed using a polynomial transformation and a bilinear interpolation. This restricts further the parameter space to be explored and is fully justified given that position angle of galaxies on the sky distribute in a random way: the position angle cannot be a principal component. Note that we do not apply any other re-scaling, e.g. of parameters such as ellipticity, which do not distribute in a random way.
	
Any companions to the galaxies used to build the PCA basis are a possible source of artefacts. Companion galaxies are frequent enough to have an important weight in the final basis. This can result in removing part of the lensed object at the end of the process or, conversely, to create fake lensed objects.  

In order to avoid this effect, we select only galaxies with no bright companions or with companions far away from the center of light. This method results of course in reducing the size of the PCA basis. To include more "companion-free" galaxies, one often has to widen the original selection function, at least in surveys of limited volume, and this may results in a PCA basis less representative of the considered sample. The selection also involves reducing the efficiency of the removal of galaxies with companions. In order to search for strong lensing around that peculiar kind of morphologies, one can devise a masking strategy, but this has not been considered in the present study.


The PCA analysis is computed by building a matrix  ${\bf X_b }$ in which each of the $n$ columns is an image from the basis set, reshaped as a vector  of size $n^2$. A singular value decomposition is performed on the covariance matrix  of the elements of the basis, ${\bf X_b }$, which  boils down to find {\bf $V$}, and {\bf $W$} verifying 
\begin{equation}
\rm	\bf X_b^{T} X_b = VWV^{T},
\end{equation}

where $\rm \bf W$ is a diagonal matrix.
The singular value decomposition of ${\bf X_b }$ is written  

\begin{equation}
\rm \bf X_b = U\Omega V^{T}, 
\end{equation}

with $\rm \bf \Omega^{2}=W $, and ${\bf U}$ the matrix of the eigenvectors for the decomposition of $\rm \bf X_b$. Therefore, the eigenvectors $\rm \bf Ei $ can be recovered from the singular value decomposition of the covariance matrix 

\begin{equation}
\rm	\bf Ei = X_{b}V^{t}W^{-1/2}. 
\end{equation}		 

 The decomposition of an $n\times n$ image of galaxy reshaped as a column vector, $\rm \bf X_{set}$ (not necessarily in the basis) can now be decomposed as
\begin{equation}
\rm	\bf \alpha_{set} =\ Ei^{T} X_{set},
\end{equation}	

where $\rm \bf \alpha_{set}$ is a ${\bf N_{gal}}$-sized vector of PCA coefficients that represents the image $\rm \bf X_{set}$. 
 
A partial reconstruction of the image is done by using only the $k$-first coefficients of the PCA, i.e. the $k$ most significant coefficients. The estimated reshaped image is
\begin{equation}
\rm	\bf \tilde{X}_{set} = \ Ei_{[0..n^2, 0..k]}\alpha_{set[0..k]}.
\end{equation} 

As the basis does not represent anything but the variations in shapes of the central parts of the galaxies, they will be the only reconstructed objects. The remaining companions 	are much less represented in the PCA basis. Rare structures such as Einstein rings or multiply imaged objects are very little represented in the PCA basis. Using a limited number of PCA coefficients during the reconstruction will therefore create images of lens galaxies without any significant lensed structure potentially present in the original data. The reconstructed PCA images can therefore be subtracted from the original data in order to unveil the lensing structures, when present. Fig.~\ref{fig:PCA_coefs} displays examples of the first PCA coefficients for the simulated Einstein rings described in Section~\ref{Euclid}.

In order to evaluate the quality of reconstruction in an objective way, we compute the reduced $\chi^2$ (per pixel) of the reconstruction in some circular area $S$ containing $N_S$ pixels:
\begin{equation}
q = \frac{1}{N_S}\sum_{i=1}^{N}\Big[ \frac{d_i-m_{i}}{\sigma_i^2}\Big] ^2
\label{quality}
\end{equation}	 
where $d_i$ are the pixels in the original image along with their photometric error $\sigma_i$, and where $m_i$ are pixel values as predicted by the PCA model/reconstruction. The radius of the circular area $S$ can be chosen to match the mean size of the galaxies in the sample.  
 
A critical step in the PCA reconstruction is the choice of the number of PCA coefficients. If all of the coefficients are used, the reconstruction will include elements of the basis that represent the noise, hence resulting in an overfitting of the data and to an apparent smoothing of the residual image obtained after subtraction of the galaxy. This can be damaging when trying to detect faints rings and arcs. Conversely, if the number of coefficients is insufficient the central galaxy will be only partially removed leaving significant and undesired structures in the residual image. 
	
In Section~\ref{Euclid}, we describe a way to choose the number of PCA coefficients in an objective way, using the reduced $\chi^2$ and we illustrate the effect of this choice using a set of simulated Einstein rings, as they would be seen with the ESA Euclid satellite \citep{Euclid}.	

\section{Finding the lensed images, arcs and rings}
\label{Finder}
Once a galaxy is removed from the image, the second step is to search for any residual lensed signal. In this paper, we focus on partial or full Einstein rings. We investigate two different approaches. The first one uses a curvelet filter \citep{starck2002}, optimized to enhance any arc-like structure, on images reshaped in a polar grid. The second method uses {\tt SExtractor} \citep{Bertin1996} to identify remaining sources in the residuals and to assess whether they are lensed images according to their orientation and elongation. 

\begin{figure}[t!]
\begin{center}
\includegraphics[width=8.3cm]{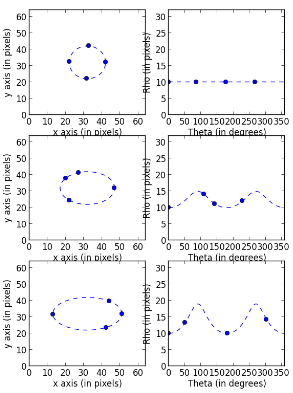}
\caption{{\it Left panels}: schematic view of rings (dashed line) and multiple images (blue dots along the ring tracks). {\it Right panels}: their corresponding transform in polar coordinates.
\label{fig:fit_model}}
\end{center}
\end{figure}

\subsection{Polar transform}

A simple way to detect full or partial rings can be devised by turning the Cartesian coordinate system of the data into the polar one. The polar coordinates $(\rho, \theta)$ are chosen so that the origin is the center of the galaxy that has been removed using the PCA decomposition. The polar-transformed image is built by creating a new grid of pixels and by asking, for each pair of $(\rho, \theta)$ coordinates, the value of the pixels in the original $(x, y)$ Cartesian grid. This involves an interpolation process giving the pixel intensities $I_{\rm pol}(\rho,\theta)$ as a function of the pixel intensities in the original image $I(x,y)$, with the standard relations $x = \rho \cos(\theta)$ and $y = \rho \sin(\theta)$.

By construction, the polar transform centered on the lens galaxy barycenter, turns a circle into a line, as illustrated in Fig. ~\ref{fig:lens_detector}. The problem of ring detection is then reduced to a problem of line detection. The polar image's columns are collapsed into a vector containing the median value of each column. If the original image contains a ring, this vector will present a spike, whose position directly gives the radius of the ring, as illustrated in Fig.~\ref{fig:spike}. In practice, we define a threshold that determines if the maximum of the vector stands for a ring or not. Figs. ~\ref{fig:lens_detector} \& \ref{fig:spike} show the different steps of the ring detection.

As the rings are not always perfectly circular but elliptical, their shape in polar coordinates can deviate significantly from a straight line, as is the case in Fig.~\ref{fig:lens_detector}. In most cases, looking for straight lines in polar coordinates is sufficient to detect rings, at least for moderate  ellipticities. However, it is possible to refine the detection criterion by fitting an ellipse in polar coordinates, 

\begin{equation}
\rho(\theta) = \frac{a b}{\sqrt{(b \cos \theta)^2  + (a \sin \theta)^2}},
\end{equation}	 

where $a$ and $b$ are the semi-major and semi-minor axes of the ellipse and where the origin of the system is centered on the lensing galaxy. In order to find point-source components superposed to the rings (or simply lensed point sources), one can add simple Gaussian profiles to the fit or the actual instrumental/atmospheric PSF. Alternatively, one can implement the detection scheme of \citet{Meneghetti2008} to find brightness fluctuations along the arcs. Different typical lensing configurations are shown to illustrate this in Fig.~\ref{fig:fit_model}.

\begin{figure*}[t!]
\begin{center}
\includegraphics[width=4.3cm, height=4.3cm]{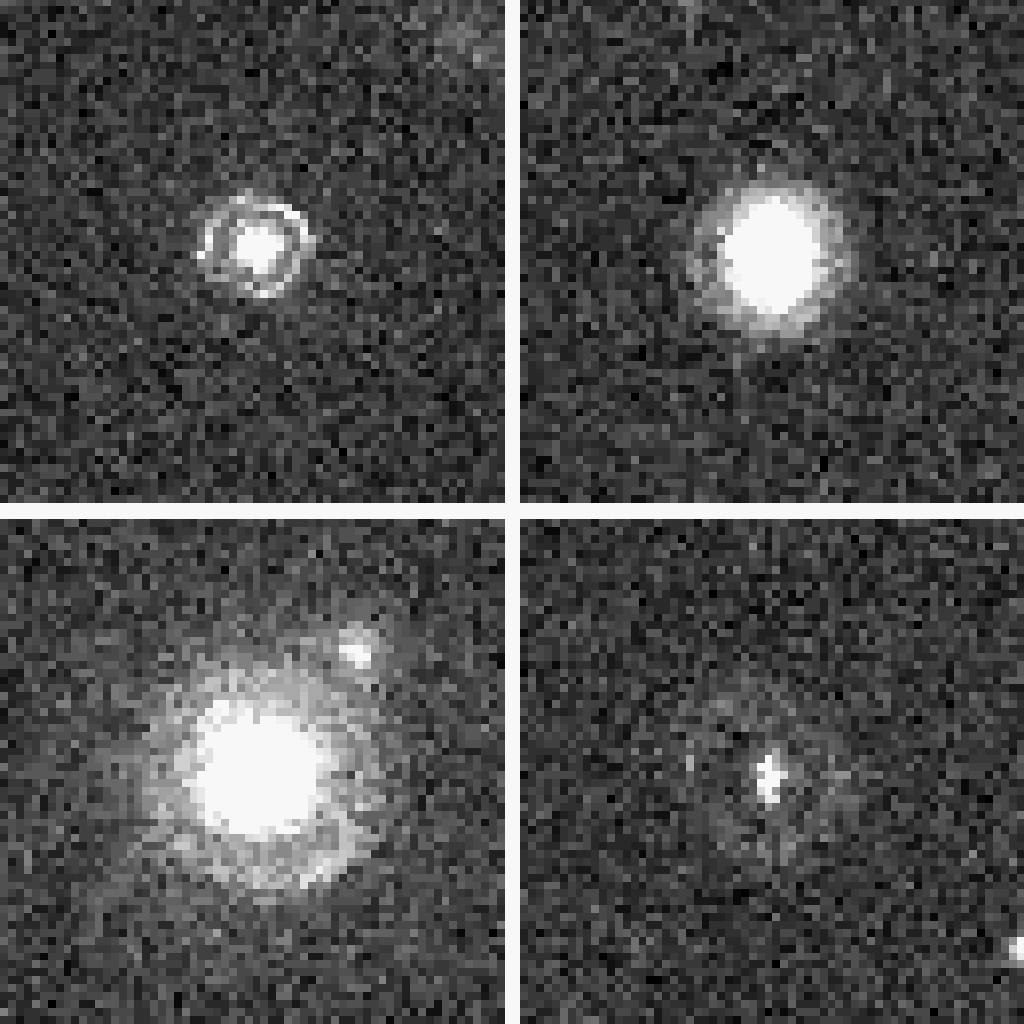}
\hskip 7pt
\includegraphics[width=4.3cm, height=4.3cm]{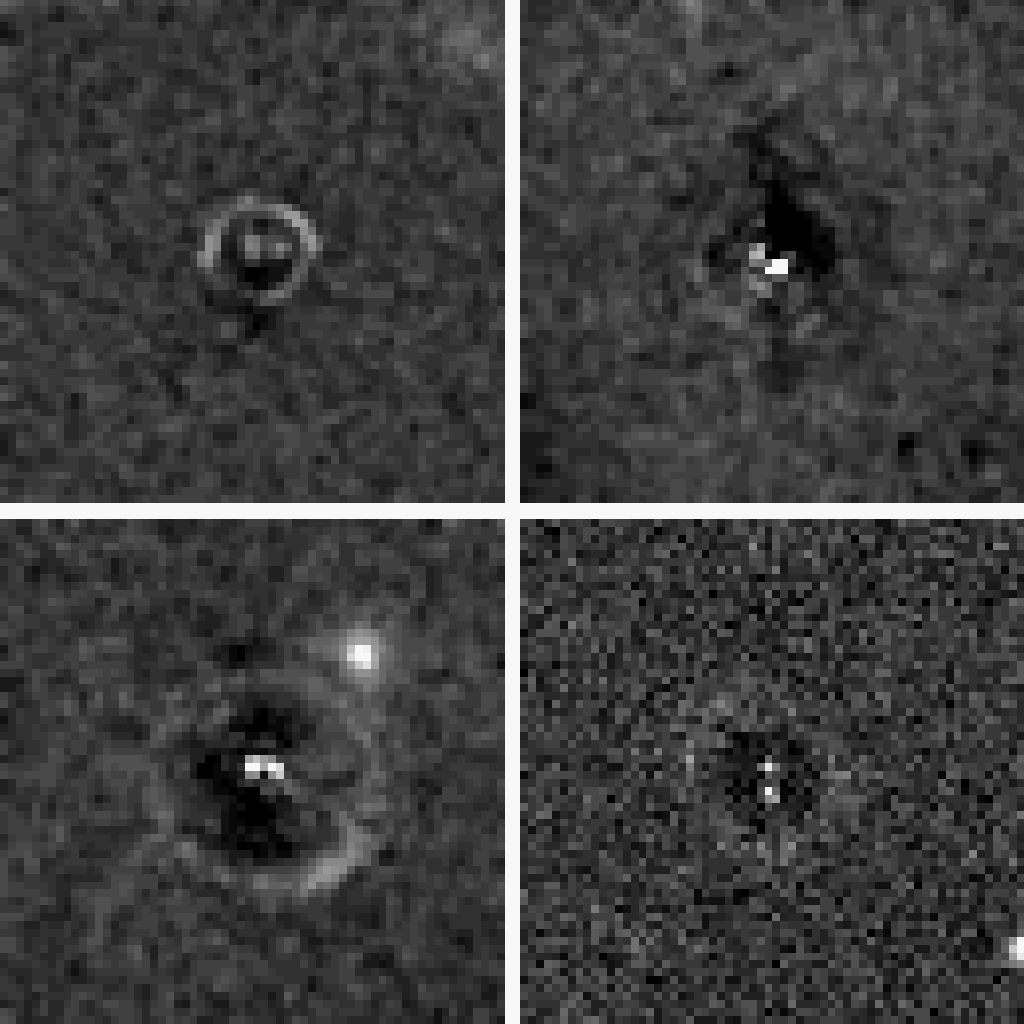}
\hskip 7pt
\includegraphics[width=4.3cm, height=4.3cm]{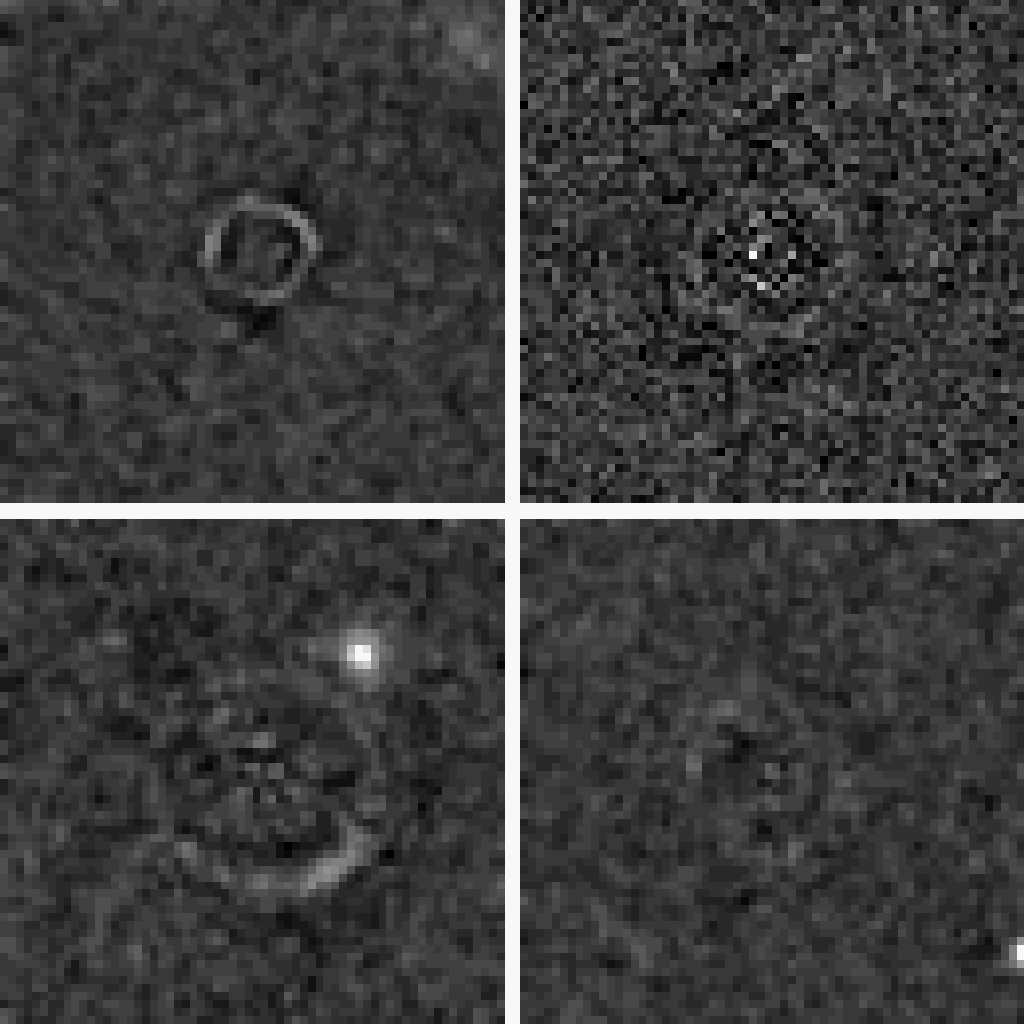}
\hskip 7pt
\includegraphics[width=4.3cm, height=4.3cm]{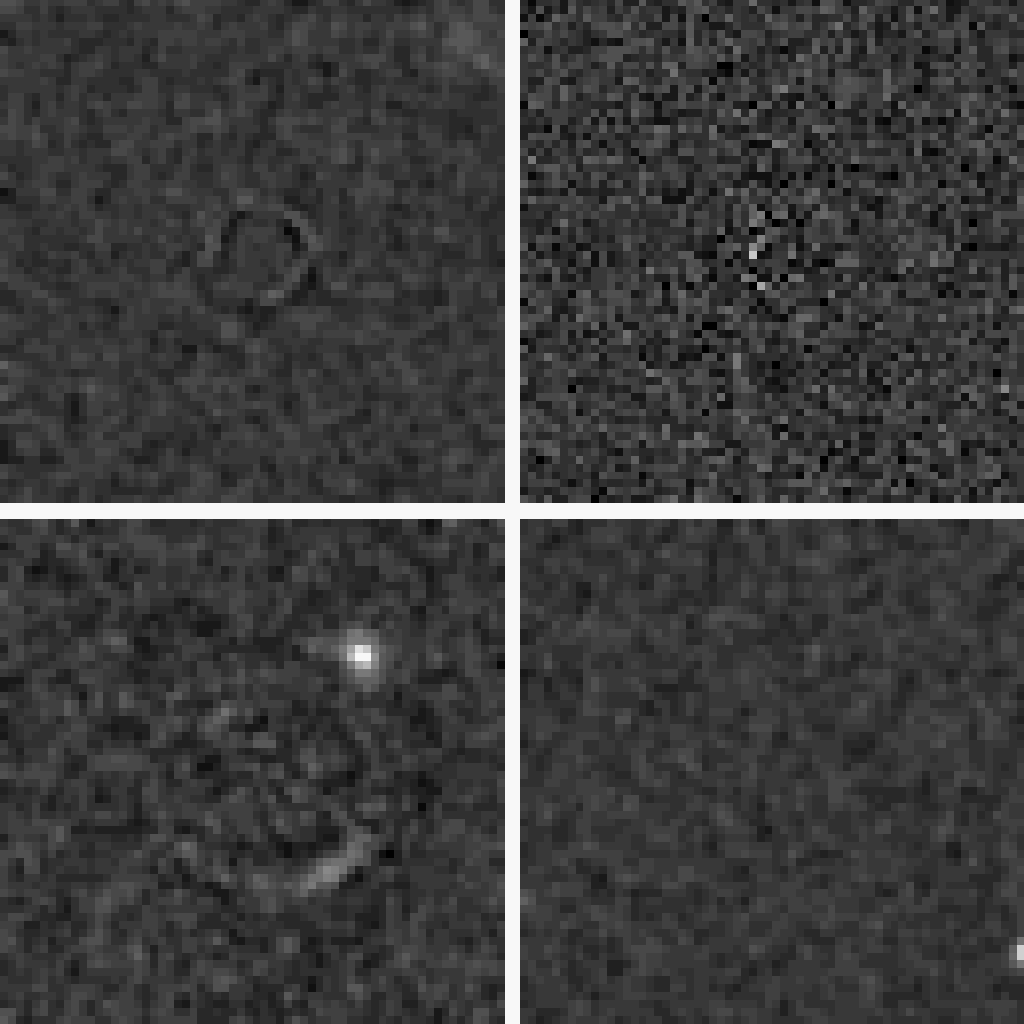}
\caption{Result of the galaxy removal on four of our simulated Einstein rings. The left hand side panel displays the four original images. From left to right, the other panels display galaxy removals when 10, 50 and 200 PCA coefficients are used. The reduced $\chi^2$ are respectively $q=1.74$, $q=1.00$ (i.e. optimal number of coefficients), and $q=0.9$. \label{fig:cleaning}}
\end{center}
\end{figure*}

\subsection{Island finding: the use of {\tt SExtractor} parameters }

An alternative method for assessing the presence of lensed structure in fields is to characterise all sources in the field, and use the measured parameters of these sources in order to identify patterns among them. This process begins with the use of {\tt SExtractor} to identify sources in the field above a signal-to-noise threshold. The flux, ellipticity, tangentiality (closeness of the position angle to 90$^{\circ}$ to a vector from the field centre to the object), and distance from the field centre are measured. In addition, flux islands (which may contain one or more {\tt SExtractor} components) are identified and the third moments of the flux distribution are measured. Third moments are sensitive to bent or arc-like structures, which are hard to detect from single components alone. For the current purpose, we define a combination of third moments $\zeta$ as:

\begin{equation}
\zeta = \frac{1}{2} \log_{10} \left[(\mu_{30}+\mu_{12})^2+(\mu_{21}+\mu_{03})^2\right]  ,
\end{equation}

where

\begin{equation}
\mu_{mn}=\sum_{n, m} d(x,y) x^m y^n
\end{equation}

where $d(x,y)$ is the data value in terms of offsets $x$ and $y$ from the brightest pixel in the island. This statistics, as a combination of third moments is sensitive to bending and is also invariant under scaling and rotation.

A Point score is then assigned to each component according to the elongation of the component and its tangential orientation with respect to the field centre. In addition, components with similar radii are weighted upwards in the point score allocation, and components which are part of an island with significant third moment are also weighted up. Specifically, the point score is given by the following procedure, using free parameters $p_i$ where necessary:

\begin{figure*}[t!]
\begin{center}
\includegraphics[width=18cm]{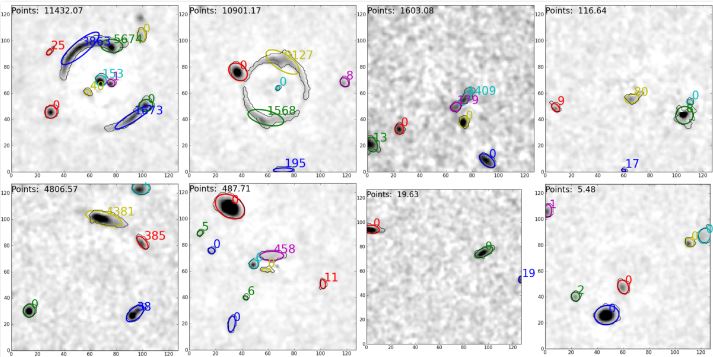}
\caption{Results of the island-finding algorithm. Each panel shows the
residual image after the PCA galaxy subtraction, with the point score
of each component given separately, and the total point score at the
top (see text). The top row shows systems which have lenses, and is ordered so
that the highest point-score is on the left and the lowest on the right.
Objects with high ellipticity and high curvature, tangential to the
radius vector from the centre of the image, are highly preferred; lens
systems without such objects are hard to recognise by eye and also tend
to attract a lower point score. The bottom row shows a sample of non-lenses,
again ordered by point score. High point-score objects are generally 
those in which chance coincidences produce configurations which mimic 
the presence of lensing.}
\end{center}
\end{figure*}
\begin{itemize}

\item Each component, unless it has a flux less than a threshold $p_{0}$, is 
assigned a point score of $10{\epsilon}^{2}exp(-t^2/p_1^2)$, where
$\epsilon\equiv a/b$ is its elongation and $t$ is the difference 
between its tangentiality and the angle tangential to the radius vector 
to the point. In general, we use Gaussian penalty functions where we wish
to select for a value close to one which would be expected for lensing,
and power laws for quantities which we wish to maximise. The $\epsilon^2$ 
dependence results from a limited amount of experimentation by hand, 
although such dependencies can ideally be optimized on a larger sample.

\item The point score of any component within a factor of $p_2$ in radius from
its neighbour is multiplied by $(1.0+N/p_3)*exp[-(r-1)^2/p_4^2]$, where $N$
is the number of points assigned to the neighbour, and $r$ is the ratio of
their distances from the centre of the field. This selection favours
multiple lensed images at the same radius, although the selection will have more
effect if the individual images are themselves elongated and tangential.

\item If a component is part of an island with third moment $\zeta>p_5$,
its point score is multiplied by $[1+(\zeta-p_5)]^2$.

\end{itemize}

The six parameters $p_i$ are then optimized on a small training set of lenses before being applied to the dataset. A variable point-score threshold can be used for lens detection, completeness generally being achieved at the expense of purity of the resulting sample. 

\section{Application to Euclid-like simulated images}
\label{Euclid}

The "lens finder" described in Sect.~\ref{PCA} is designed to process large imaging data sets. Although the pre-selection of the galaxies to be searched for lensing may require color information, the new algorithm proposed in this paper can be applied to single-band data to perform a purely morphological search. In the following, we evaluate the performances of the method using simulated images of Einstein rings, as they would be seen with the ESA Euclid satellite \citep{Euclid}. 

The image simulations are provided through the Bologna Lens Factory (BLF) project\footnote{www.bolognalensfactory.wordpress.com}.  This is a project dedicated to performing lensing simulations and providing realistic mock data for a large variety of lensing studies from large scale weak lensing, to galaxy cluster lensing and strongly lensed quasars.  For the purposes of this work, images were created to specifically mimic the expected Euclid images in the visible instrument, as described in \cite{Euclid}. The pixel size is 0.1\arcsec\ and the PSF is Gaussian with a Full-Width-Half-Maximum (FWHM) of 0.18\arcsec. The surface brightness is translated into photon counts taking into account the expected instrumental throughput in the VIS band. Background counts from zodiacal light are added, assuming a brightness equal to 22.8 $mag/arcsec^2$. Noise is then calculated taking care of Poisson statistics, flat-field error and read-out \citep{Meneghetti2008}. 
The lensing and image construction is done with the GLAMER lensing code \citep{M&P2013,PM&G2013}.  The pre-lensed galaxy surface brightness models and mass distribution are provided by the Millennium Run Observatory \citep[MRObs;][]{2013MNRAS.428..778O}.  Each galaxy is represented by a bulge and a disk component whose properties are predicted by a semi-analytic galaxy evolution model.  The mass distribution consists of halos identified in the Millennium Nbody simulation.

The lensing simulations were done as follows.  The halos in the catalog are represented by NFW halos \citep{1997ApJ...490..493N} with Singular Isothermal Ellipsoids (SIEs) in their centers to represent the baryonic galaxy.  This model has been shown to fit observed Einstein rings well \citep{2007ApJ...667..176G}.  The NFW profile is fit to the mass and peak circular velocity of the halo found in the Millennium simulation.  The mass and velocity dispersion of the SIE component is set by the stellar mass to halo mass relation of \cite{2010ApJ...710..903M} and the Faber-Jackson relation \citep{1976ApJ...204..668F}.  The lensed image of every source within a 0.1~deg$^2$ light cone down to 28th magnitude in I band is constructed and put into a master image.  This image contains only a few strongly lensed objects because the source density is small enough that it is rare to have a visible object within a caustic.  To boost the number of strong galaxy-galaxy lenses, all the critical curves and their associated caustics in the field are found for a source redshift of $z_s=2.5$ and a source galaxy is moved to be near the caustic.  The sources are taken randomly from galaxies within the light cone at a si\-mi\-lar redshift.  Then the lensed image of this source is constructed and added to a  $200 \times 200$~pixel cutout stamp from the master image.  Images with and without the added source are provided and an image with only the added, lensed source are provided.  All images are provided with and without the noise and PSF effects.  A catalog of all the critical curves and caustics is also provided with their locations and properties such as average radius and area.

Since we are not concerned with predicting the statistical properties of the lenses in this paper, many of the precise details of these simulations are not important (for example the distribution of source and lens redshifts, morphologies, luminosities, etc.).  The performance of the PCA lens finder will be stated in terms of the signal-to-noise ratio of the Einstein ring so the simulations are only required to represent the variety of expected lenses and not their precise distribution.

The set of Euclid simulation images consists of 3000 galaxies with a full or partial background Einstein ring and of a training set of 1250 galaxies with no lensing. Adding more galaxies to the training set does not change significantly the PCA basis. Among the 1250 non-lensing galaxies of the training set, 1000 are used to build the PCA basis in order to search for lensing in the 3250 images, 3000 of which containing Einstein rings. Note that with real data, the training set can be the whole data set itself, as galaxies with lensing features are rare. 

Building the PCA basis for the 1000 Euclid galaxies, which are 128 pixels on-a-side, takes about 40 minutes on a single processor. Using this PCA basis, doing the galaxy reconstruction and subtraction takes less than a minute more for the whole data set, i.e. 3250 images. In terms of {\tt cpu}, the PCA method is therefore well tractable and applicable to large data sets.

\subsection{Quality of the central galaxy reconstruction}

The quality of the PCA reconstruction depends on 3 main factors: 1- the range in galaxy sizes, 2- the presence of companions near the galaxies used to build the PCA basis, 3- the number of PCA coefficients to be used.  

In order to minimize the parameter space to explore, all galaxies are first centred on the central pixel of the FITS stamp and rotated so that their long-axis aligns with the image rows. If necessary, the resulting images are zero-padded and trimmed to a common size. In the present case we use $128 \times 128$ pixels. 

In order to minimize  the contamination of the PCA basis by companions to the galaxies in our sample, we only select the stamps that have no companion brighter than 50\% of the maximum brightness of the main galaxy in a range of less than 10 pixels to the patch's center, i.e. 1\arcsec\ given the Euclid pixel size of 0.1\arcsec. 

To estimate the number of PCA components, we carry out different reconstructions with an increasing number of PCA coefficients. We stop adding coefficients when reaching an acceptable quality, i.e. when there is no residual above the noise level. A good reduced $\chi^2$ is when $q$, (Eq.~\ref{quality}) remains between 1 and 1.5, i.e. when the mean $\chi^2$ per pixel is on average close to $1\sigma$. Indeed, if the pixels in the residuals are highly correlated due to a reconstruction that includes coefficients representative of the noise, the reduced $\chi^2$ becomes smaller than 1. Conversely, when the residuals contain important patterns due to an insufficient reconstruction, $q$ is significantly larger than 1. This is illustrated in Figs.~\ref{fig:q_ncoeffs} \& \ref{fig:quality_density} for the specific case of our Euclid simulation, where a good reconstruction is achieved for a number of PCA coefficients of about 50, i.e. the minimum number of coefficients required to reach $q\sim 1$.

\subsection{The effect of galaxy sizes}

Even for relatively smooth light distributions, like early type galaxies, a careful balance must be found between the number of galaxies in the training set and the range in galaxy sizes. We investigate in the following the influence of the distribution of the galaxies in sizes for the specific case of our Euclid simulations. 

\begin{figure}[t!]
\begin{center}
\begin{tabular}{p{0.1cm} c}
\begin{turn}{90}{\tt $\qquad$ $\qquad$ $\quad$ Reduced $\chi^2$ (q)} \end{turn} &
{\includegraphics[trim=2cm 1.3cm 0cm 0cm, clip=true, width=8.5cm]{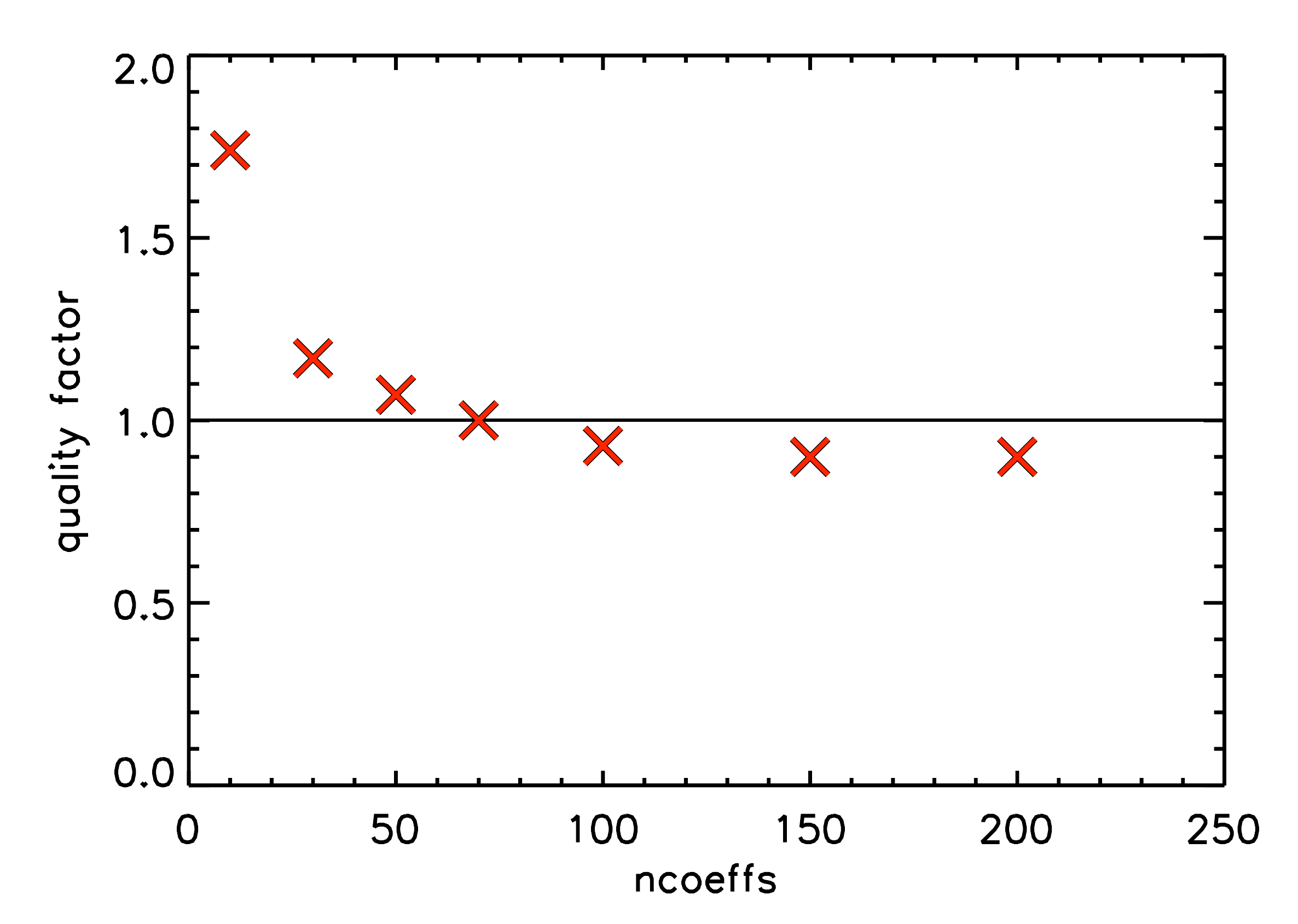}} \\
 & {\tt  number of coefficients}
\end{tabular}
\caption{Reduced $\chi^2$, as a function of the number of coefficients used in the reconstruction. Only 50-70 coefficients are needed to reach a reduced $\chi^2$ of $q\sim 1$ in the case of our Euclid simulations. \label{fig:q_ncoeffs}} 
\end{center}
\end{figure}

To do so, we bin the sample in galaxy sizes, keeping 100 galaxies per bin and we build the PCA basis for each bin of size, i.e. like in Fig.~\ref{fig:q_ncoeffs}. Note that rescaling the galaxies in $R_{eff}$ is also an alternative, but we try as much as we can to avoid alter the data before building the PCA basis. Rescaling in $R_{eff}$ may be considered for small samples of galaxies that cannot be binned in galaxy size. The images are then reconstructed using different number of coefficients. The quality of reconstruction, estimated using the median $q$ factor over all images of the sub-sample, is then evaluated. Fig.~\ref{fig:q_ncoeffs} suggests that 50-70 coefficients is an optimal number to reach a reduced $\chi^2$ close to 1.

\begin{figure}[t!]
\begin{center}
\begin{tabular}{p{0.25cm} c}
\begin{turn}{90}{\tt $\qquad$ $\qquad$ $\quad$ Reduced $\chi^2$ (q)} \end{turn} &
\includegraphics[trim=2cm 1.3cm 0cm 0cm, clip=true,width=8.2cm]{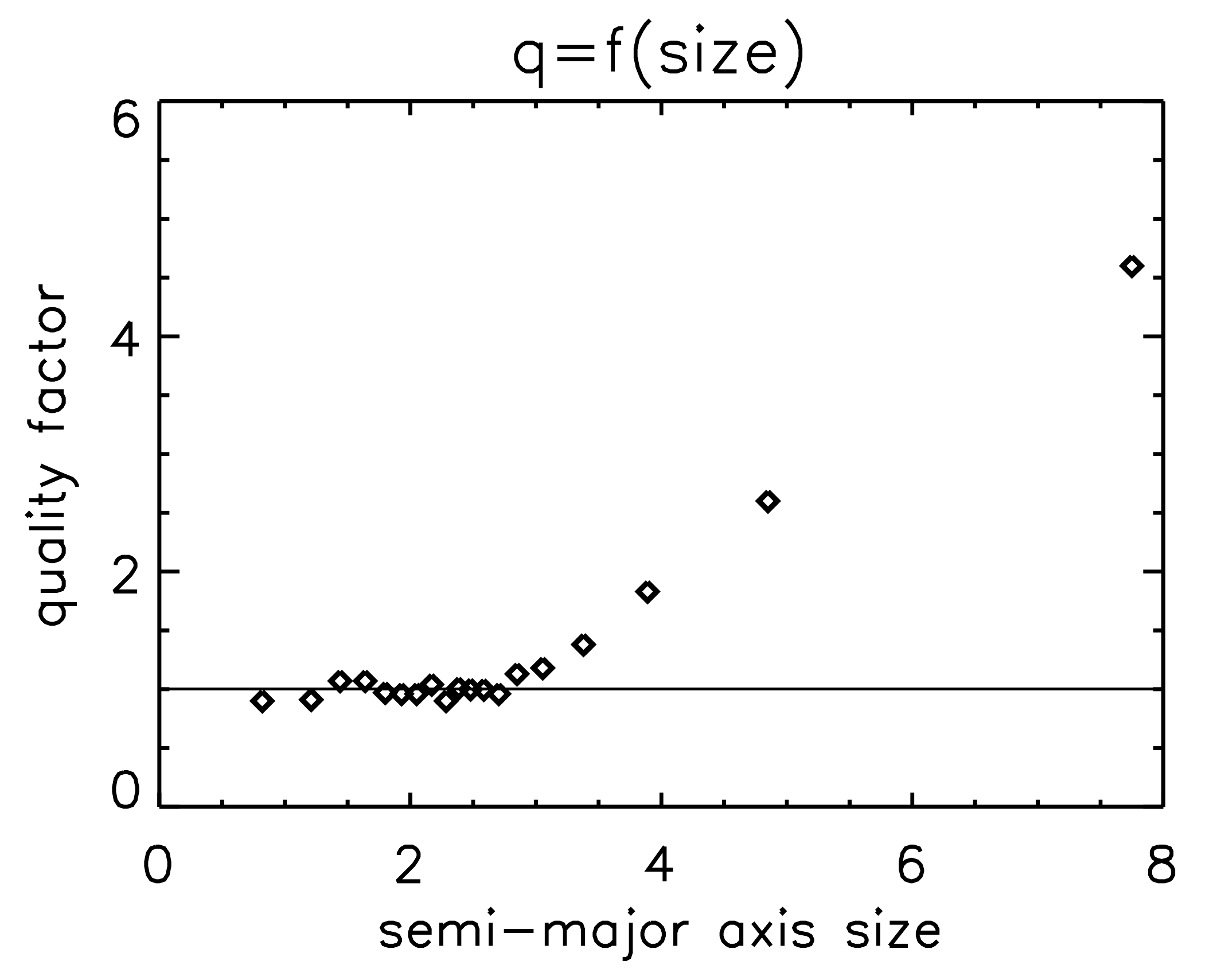} \\
 & {\tt  galaxy size in pixels along the semi-major axis}
\end{tabular}
\caption{Quality of the reconstruction of the simulated Euclid lenses as a function of the average size of the galaxies in pixels, as measured with {\tt SExtractor}. The pixel size of the images matches that of Euclid, i.e. 0.1\arcsec. As big galaxies are rare, they are less well represented in the PCA basis and they are therefore less well modeled. 
\label{fig:quality_density}}
\end{center}
\end{figure}

Fig.~\ref{fig:quality_density} shows how $q$ rises when galaxies are getting bigger than a semi-major axis bigger than 3 pixels. As big galaxies are less represented in the PCA basis, because of their scarcity, their reconstruction is less accurate, hence leading to a larger $\chi^2$.

It is therefore very important to carefully select the range of size that we want to investigate when building the PCA basis and to ensure that a sufficient number of galaxies are available to represent the full variety of structures in the sample/bin. Indeed, for bigger galaxies, where Einstein rings are more likely to be found, the number of objects contributing to the basis is reduced, simply because big galaxies are rare. 




\subsection{Completeness and purity}

\begin{figure*}[t!]
\hskip -10pt
\begin{tabular}{c c c}
\includegraphics[trim=1.7cm 2cm 4cm 2cm, clip=true,width=5.9cm, angle = 180]{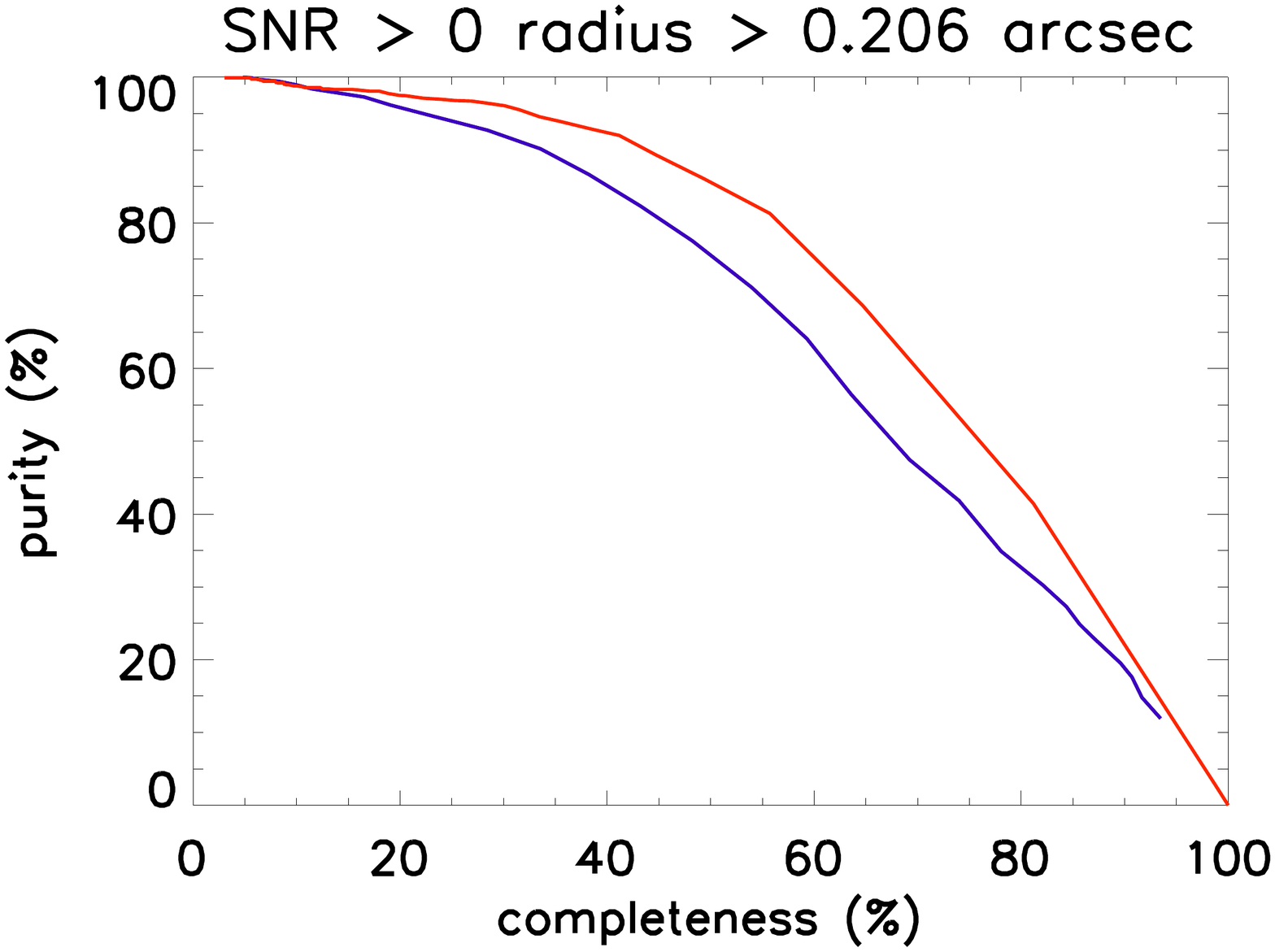} & 
\includegraphics[trim=2cm 2cm 4cm 2cm,width=5.9cm, angle = 180]{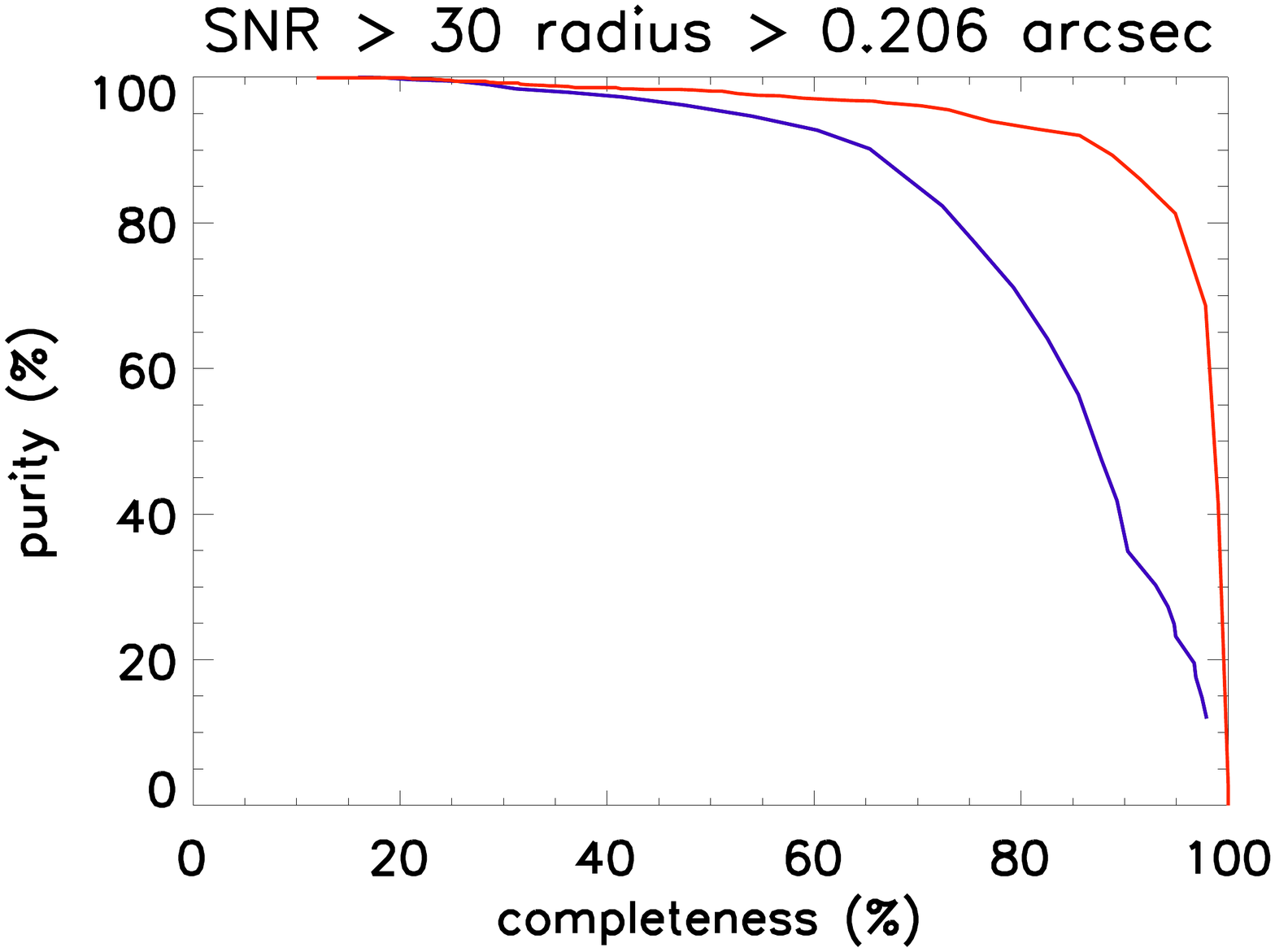} &
\includegraphics[trim=2cm 2cm 4cm 2cm,width=5.9cm, angle = 180]{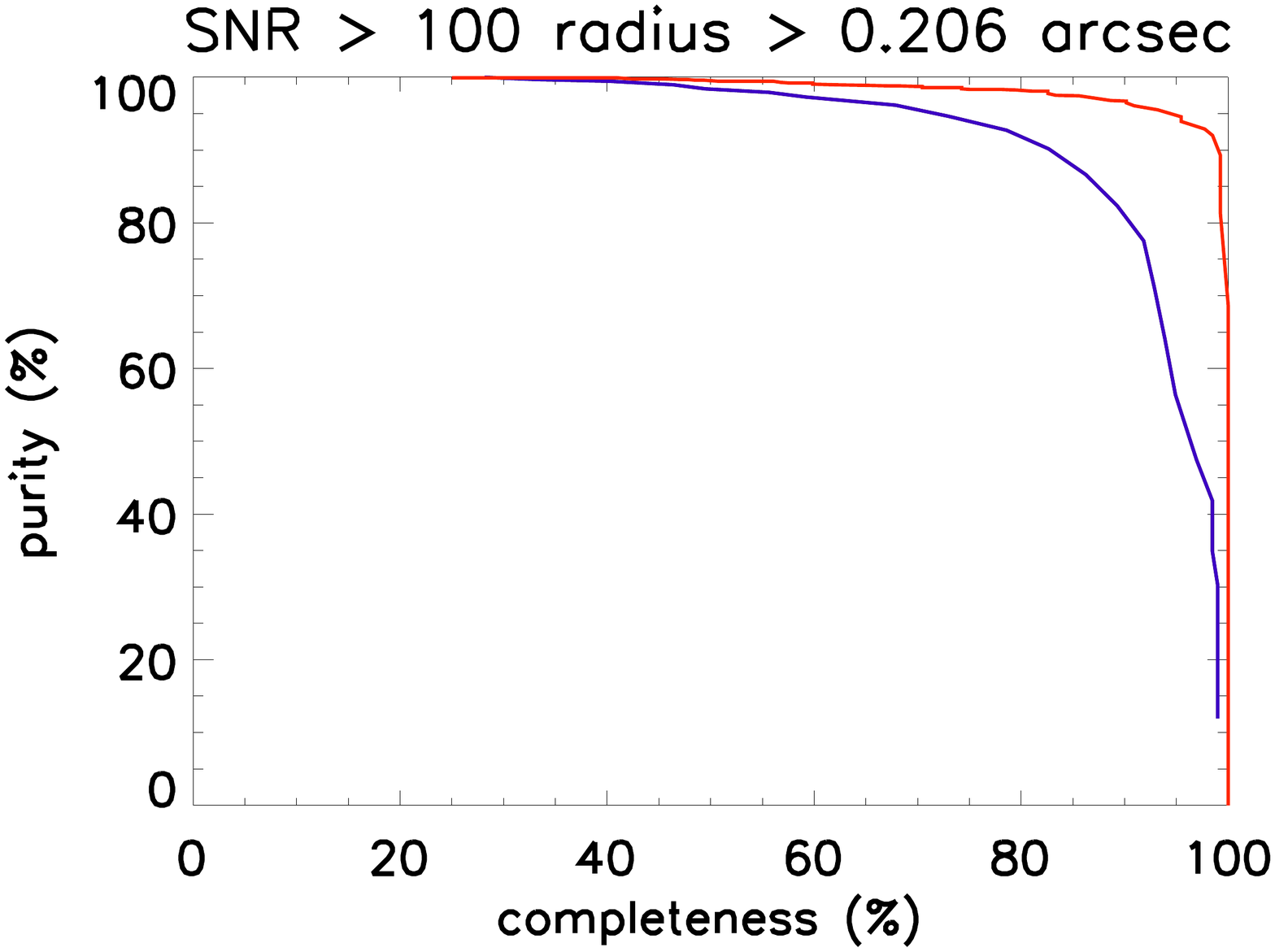} \\
\includegraphics[trim=2cm 2cm 4cm 2cm,width=5.9cm, angle = 180]{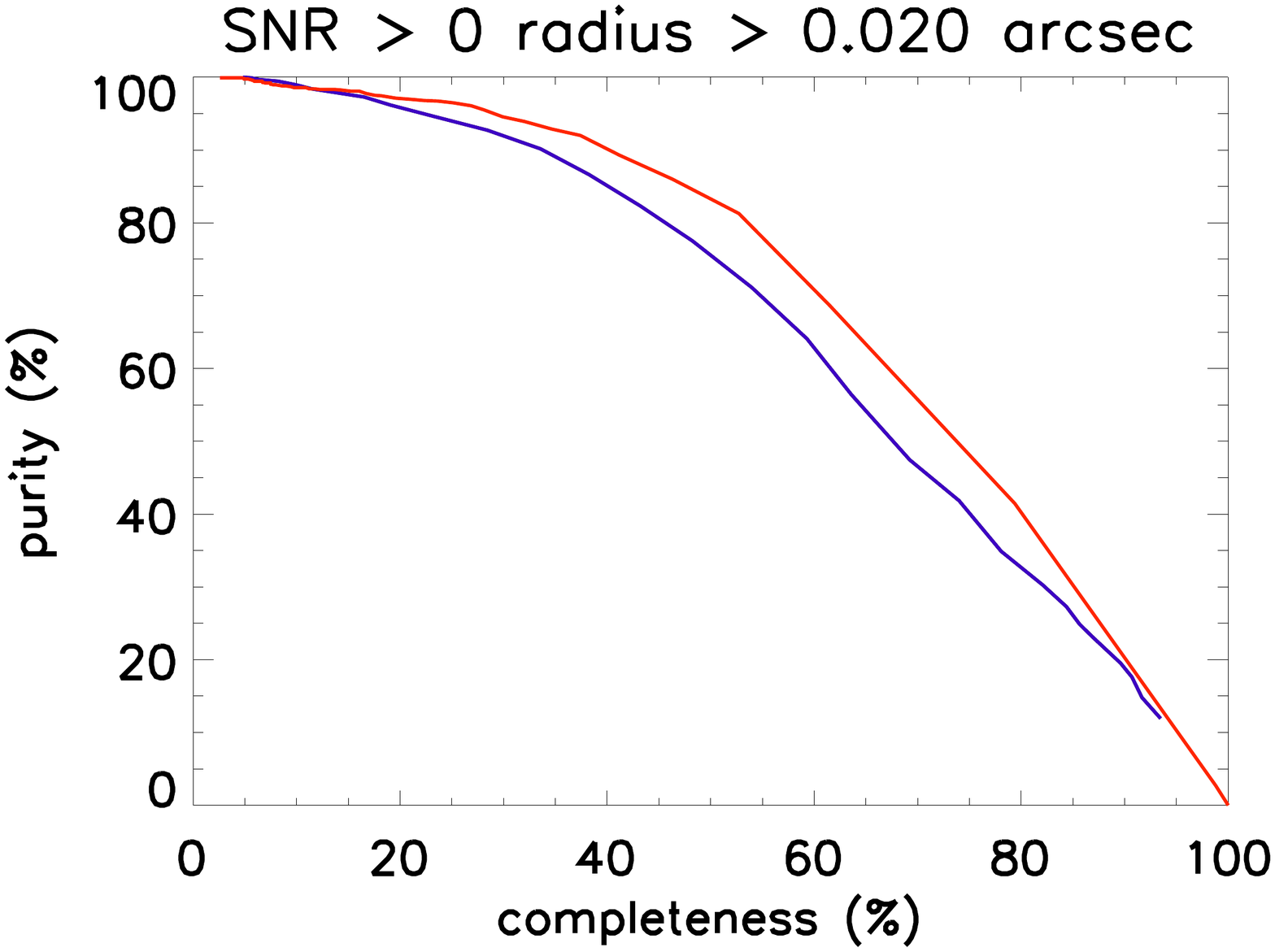} & 
\includegraphics[trim=2cm 2cm 4cm 2cm,width=5.9cm, angle = 180]{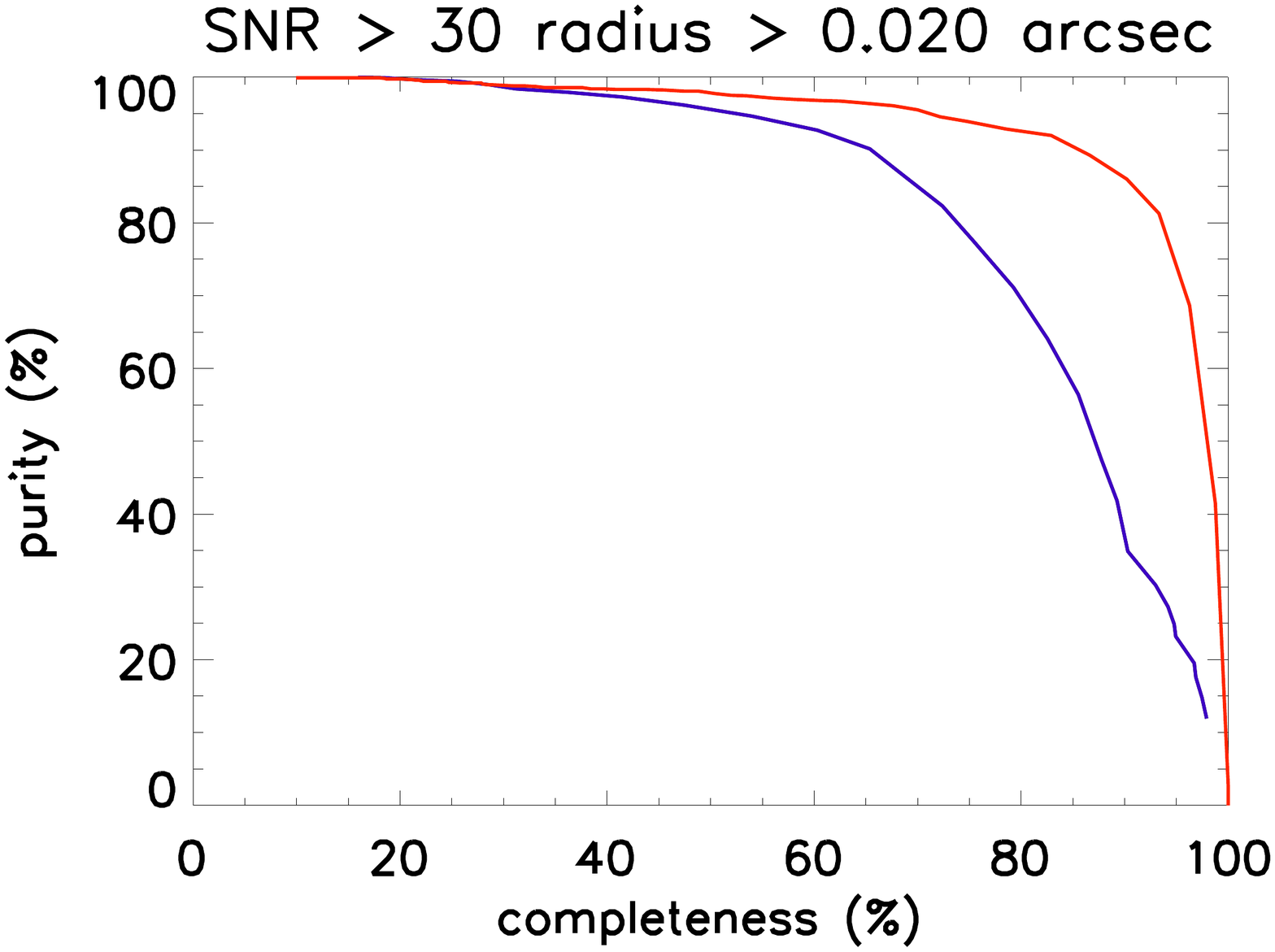} &
\includegraphics[trim=2cm 2cm 4cm 2cm,width=5.9cm, angle = 180]{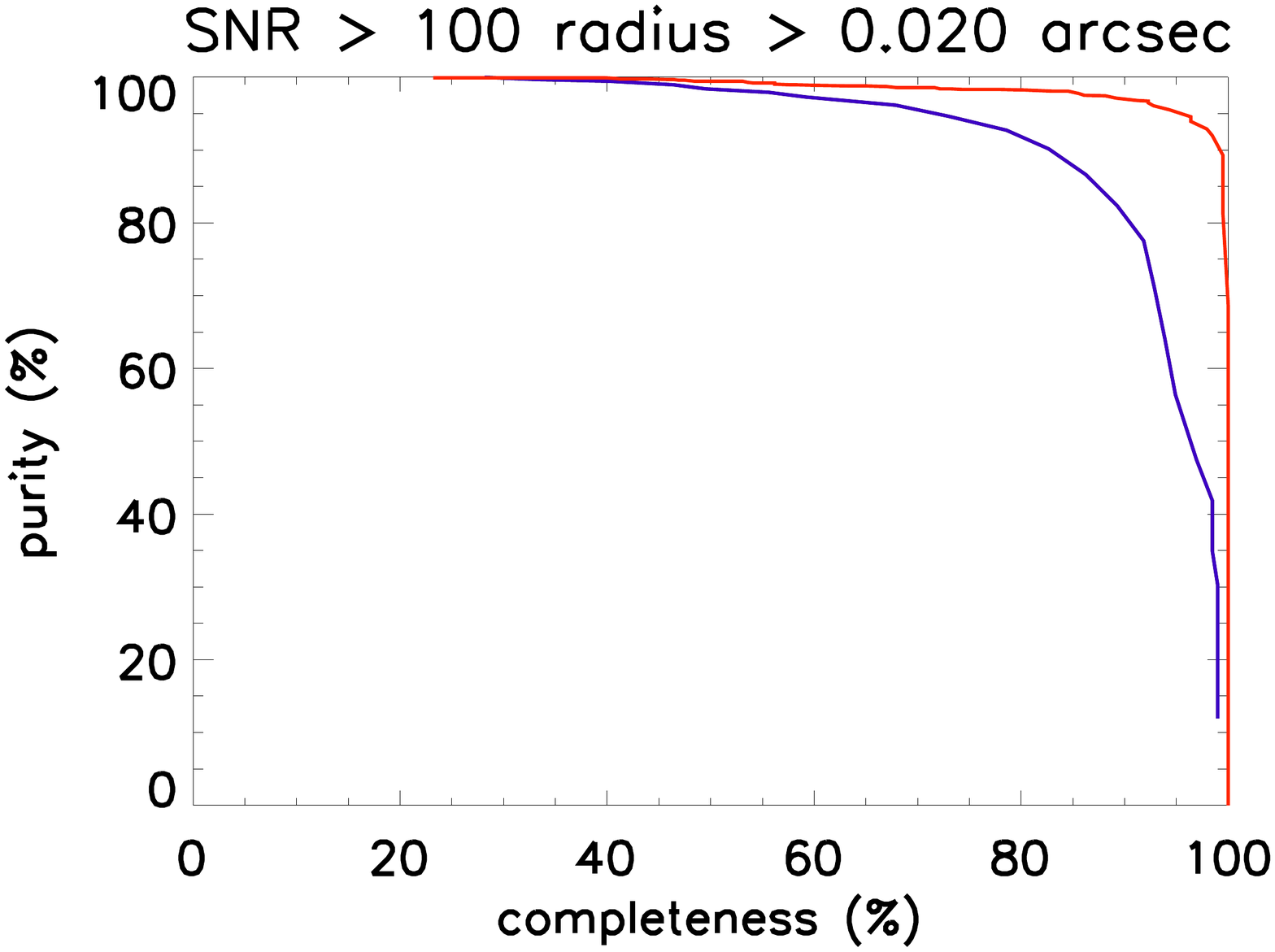} 
\end{tabular}
\caption{ Completeness as a function of purity for different thresholds of Einstein radii (expressed in terms of critical curve here) and signal-to-noise ratio with the two methods described in Sect.~\ref{Finder}: polar transform (in red) and island finding (in blue). The minimal radius in the sample is $r=0.02$\arcsec, which means that the top left panel shows the results over the whole sample.
\label{fig:Completeness}}
\end{figure*}

In order to evaluate the efficiency of the algorithm, we perform tests of detection on simulated images for which the signal-to-noise ratio and the caustic radius of the lensing galaxies are known. For this study we use a set of 3000 simulated full rings from the BLF. With these realistic Euclid-like ring images and the associated noise images we can compute the SNR for each Einstein ring:
\begin{equation}
	SNR = \frac{S}{\sigma\sqrt{N_i} },
\label{SNR}
\end{equation}

where $N_i$ is  the number of non-zero pixels in the noise free ring image, $\sigma$ is the rms noise per pixel and $S$ is the total flux in the ring. The analysis of the simulated images is done by building a PCA basis using 1000 galaxies from a set of non lensing galaxies. The detection algorithms, described in Section~\ref{Finder} are then applied to the 3000 images with lensing and to the 250 images without lensing . The island finding algorithm has been trained on a set of 167 images of lensed rings provided by the BLF, together with another set of 200 images which did not contain lenses. The parameters were optimized here, and then re-optimized on the dataset itself. The output of the process is compared with the known answer from the simulations to evaluate the completeness and the purity of the derived lens catalogues. 

As the fraction of non-lens images in the sample is small compared to reality, we rather define the purity as the fraction of non-lens images that have not been detected instead of the fraction of true positive among all the detected lensed images:
\begin{equation}
{\rm Purity}  =  1-\frac{N_{\rm false\ positive}}{N_{\rm false\ positive} + N_{\rm true\ negative}}.  \label{equation:purity}
\end{equation}

The completeness is expressed as the fraction of actual lens images that have been detected over the whole sample of lenses:
\begin{equation}
{\rm Compl.}  =  \frac{N_{\rm true\ positive}}{N_{\rm true\ positive} + N_{\rm false\ positive}}.\label{equation:completeness}
\end{equation}

Fig.~\ref{fig:Completeness} shows the purity as a function of completeness for both methods. Different thresholds in signal-to-noise ratio and critical curve for the lensing have been considered. Although both methods are comparable at low completeness, at high completeness levels the {\tt SExtractor} algorithm generally leads to lower purity, corresponding to more false positives. This problem appears worse at high signal-to-noise levels, because the number of false positive detections in the non-lens sample remains constant while the number of true positives declines. This is likely to be due to the attempt to preserve at least some sensitivity to only marginally extended components, corresponding for example to quadruply imaged sources of modest extent. The algorithm is therefore more vulnerable to chance alignments between external components; work is under way to alleviate this problem, and particularly to use colour information to distinguish between genuine and chance alignments. In the context of the present work, we stick to single-band detections. The results tend to show that we can detect rings almost independently on the radius. For instance, with the polar transform method and a signal-to-noise ratio higher than 30, one can reach a completeness of 90\% for a purity of 86\%.

\begin{figure}[p!]
\begin{center}
\includegraphics[width=1.9cm]{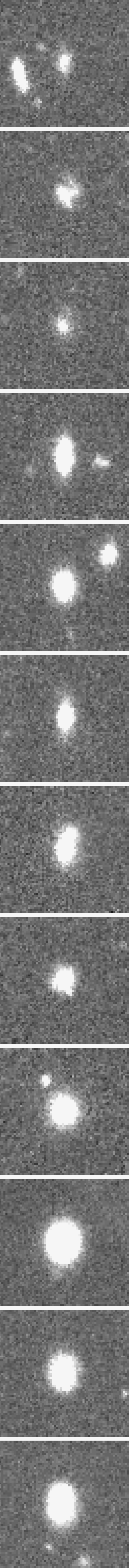}
\includegraphics[trim=9.7cm 0.02cm 9cm 0cm, clip=true,width=1.798cm]{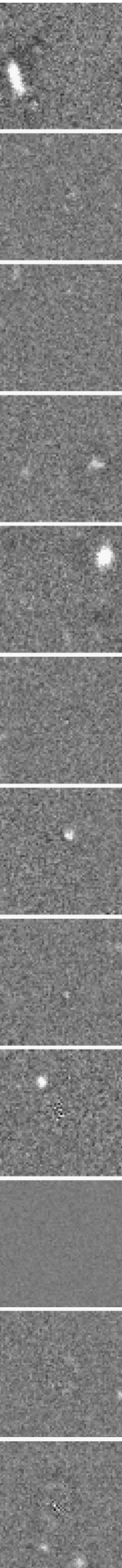}
\includegraphics[width=1.9cm]{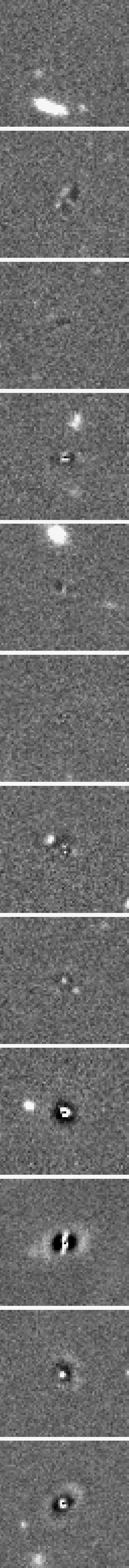}
\includegraphics[width=1.9cm]{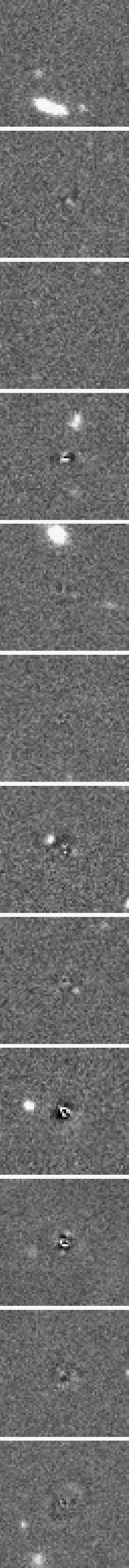}
\caption{Comparison of different galaxy-removal schemes applied to deep CFHT images. The first columns shows the original image. The second shows the residual image after subtraction of a PCA reconstruction of the galaxy. The third and fourth columns show the subtraction of a single and double elliptical sersic profile respectively, using {\tt GALFIT3}. Note that the PCA-subtracted images are rotated by construction of the PCA basis but the {\tt GALFIT}-subtracted images are not, in order to avoid interpolation when not mandatory. 
\label{fig:real_data_subtraction}}
\end{center}
\end{figure}

\section{Application to real data}
\label{Stripe82}

In the above, we test our lens finder on simulated images that mimic Euclid images in the VIS band. An obvious question is whether the algorithm performs in a satisfactory way on real data. While carrying out a ring search on a large data set is outside the scope of this paper, we can nevertheless test how our PCA decomposition of galaxies compares with other more traditional ways of removing lensing galaxies. 

In order to do that, we use the deep and sharp optical images taken with MEGACAM at the CFHT to map SDSS stripe 82. Following the same procedure as with the Euclid simulations, we set the optimal number of PCA coefficients by checking that we can actually reach reduced $\chi^2$ $1 < q < 1.2$ depending on the seeing and on the physical size of the galaxies we want to subtract. 

In Fig.~\ref{fig:real_data_subtraction}, we compare our galaxy subtraction with that done in other lens searches using single or double Sersic profiles \citep[e.g.][]{Vegetti2012, Lagattuta2010}. Not surprisingly, the subtraction with Sersic profiles performs rather well with low SNR galaxies or with small galaxies, but leaves significant residuals for large galaxy sizes. As these residuals often take the shape of a ring, they may lead to large numbers of false positives in a ring search. 

The experiment we carry out here with real data uses only 1 single field of the CFHT data of stripe 82, i.e. 1 square degree out of the 180 available. This means that the PCA decomposition uses only a limited number of large galaxies. As a consequence, using the whole 180 fields has the potential to improve further the galaxy subtraction, while profile fitting will always be limited to the information in one single galaxy and does not benefit from the global information on the shape of galaxies from a whole data set. In other words, increasing the survey size, not only increases the number of potential lenses, but also increases the density of galaxies per bin of size, hence improving the quality of the PCA basis.

\section{Conclusion}
\label{conclusion}

The two lens finder algorithms developed here all rely on a good subtraction of lensing galaxies with machine learning methods; different ideas for ring detection then allow objects with different properties to be detected on the residual images:

\begin{itemize}

\item The polar transform method enhances the signal in the residual image by applying curvelet denoising and uses a polar transform of the images to turn the problem of a circle detection to a line detection. It is designed to detect full or partial rings with or without ellipticity. 

\item The "Island finding algorithm" uses {\tt SExtractor} to detect structures in the PCA-subtracted images and to determine whether they correspond to lensed sources according to their elongation, orientation and bending. This algorithm is expected to be more efficient in finding partial arcs and multiple images.

\end{itemize}

The method is successfully applied to Euclid-like simulations. With the polar transform method, a completeness of 90\% is reached for data where the signal-to-noise in the Einstein ring is at least 30. The same simulations show that the purity of the derived ring sample reaches 86\% of the non lensed galaxies detected as false positives. 

The galaxy subtraction algorithm occurs to be efficient when applied to real data as well: our tests with CFHT images of SDSS Stripe 82 surpasses in quality the subtraction obtained with direct model fitting. 

In future work, ways to increase the purity of the algorithms will be investigated by using adapted dictionary learning \citep[e.g.][]{Beckouche2013} for galaxy subtraction. The strength of those machine learning methods should allow us to build bases adapted to more complicated problems, such as the subtraction of galaxies in clusters to detect rings produced by multiple galaxies. Better morphological selection based on PCA "clustering" or beamlet analysis \citep[e.g.][]{Donoho&Huo2002} can be used to discriminate ring-like shapes, to classify rings and arcs and to carry out galaxy classification in general, as has been done in the past with quasar spectra \citep{Boroson2010} and, more recently, with galaxy multi-band photometry \citep{Wild2014}.

\begin{acknowledgements} The authors would like to thank R. Cabanac, A. Fritz, R. Gavazzi, F. Lanusse, P. Marshall, J.-L. Starck and A. Tramacere for helpful discussions on various aspects of this paper.  
This work is supported by the Swiss National Science Foundation (SNSF). G. Lemson is supported by Advanced Grant n. 246797 ‘GALFORMOD’ from the European Research Council. B. Metcalf, F. Bellagamba, C. Giocoli and M. Petkova's research is part  of the project GLENCO, funded under the European  Seventh  Framework  Programme,  Ideas,  Grant  Agreement  n. 259349. P. Hartley is supported by a
Science \& Technology Facilities Council (STFC) studentship. 
J.-P. Kneib is supported by the European Research Council (ERC) advanced grant “Light on the Dark” (LIDA).
\end{acknowledgements}

\bibliographystyle{aa} 
\bibliography{biblio} 
\end{document}